\documentclass[journal=aanmf6,manuscript=article,layout=twocolumn,citeautoscript=false]{achemso}
\usepackage{lipsum}
\usepackage{kantlipsum}
\usepackage{verbatim}
\usepackage{sidecap}
\usepackage[mathlines]{lineno}
\usepackage{amsmath,amssymb,amsfonts}
\usepackage[osf]{newpxtext}
\usepackage{newpxmath}
\usepackage[utf8]{inputenx}
\usepackage{graphicx}
\usepackage[dvipsnames]{xcolor}
\usepackage[pdftex]{hyperref}
\setkeys{acs}{doi = true, articletitle = true}
\usepackage{braket}
\usepackage{xspace}
\usepackage{dblfloatfix}
\usepackage{soul}
\usepackage{chemformula} % Formula subscripts using \ch{}
\usepackage[T1]{fontenc} % Use modern font encodings
\usepackage{siunitx}
\usepackage{float}
\usepackage[normalem]{ulem} %needed to compile certain diff files

\DeclareMathAlphabet{\mathcal}{OMS}{cmsy}{m}{n}

\hypersetup{
	pdfauthor={Philipp John},
	bookmarksnumbered=true,
	pdftitle={Strain-free AlScN NWs },
	colorlinks,
	citecolor=black,
	linkcolor=black,
	urlcolor=black}

\author{Adriano Notarangelo}
\affiliation
{Paul-Drude-Institut f\"ur Festk\"orperelektronik, Leibniz-Institut im Forschungsverbund Berlin e.V., Hausvogteiplatz 5-7, 10117 Berlin, Germany}
\author{Rudeesun Songmuang}
\affiliation{Universit\'e Grenoble Alpes, CNRS, Grenoble INP, Institut N\'eel, 38000 Grenoble, France}
\author{Mostafa Saleh}
\affiliation{Universit\'e Grenoble Alpes, CNRS, Grenoble INP, Institut N\'eel, 38000 Grenoble, France}
\author{Natthawadi Buatip}
\affiliation{Universit\'e Grenoble Alpes, CNRS, Grenoble INP, Institut N\'eel, 38000 Grenoble, France}
\author{Bruno Fernandez}
\affiliation{Universit\'e Grenoble Alpes, CNRS, Grenoble INP, Institut N\'eel, 38000 Grenoble, France}
\author{Ileana Florea}
\author{Philippe Venn\'egu\`es}
\affiliation{Universit\'e Côte d’Azur, CRHEA, CNRS, 06905 Sophia-Antipolis Cedex, France}
\author{Aidan F. Campbell}
\affiliation
{Paul-Drude-Institut f\"ur Festk\"orperelektronik, Leibniz-Institut im Forschungsverbund Berlin e.V., Hausvogteiplatz 5-7, 10117 Berlin, Germany}
\author{Hans Tornatzky}
\author{Jonas L\"ahnemann}
\author{Thomas Auzelle}
\author{Lutz Geelhaar}
\author{Oliver Brandt}
\author{Philipp M. John}
\affiliation
{Paul-Drude-Institut f\"ur Festk\"orperelektronik, Leibniz-Institut im Forschungsverbund Berlin e.V., Hausvogteiplatz 5-7, 10117 Berlin, Germany}
\email{john@pdi-berlin.de}

\title{Molecular Beam Epitaxy of Al$_{\boldsymbol{\mathsf{1-x}}}$Sc$_{\boldsymbol{\mathsf{x}}}$N Nanowires: Towards Group-III Nitride Nanogenerators with Enhanced Piezoelectric Response}

\keywords{ScAlN, nanowires, piezoelectric energy harvesting, phase separation, effective dielectric permittivity, transmission electron microscopy, AlScN, polarity\\}

\newcommand{\AlScN}{Al$\mathrm{_{1-x}}$Sc$\mathrm{_{x}}$N\xspace}

\begin{document}

	%%%%%%%%%%TABLE OF CONTENTS%%%%%%%%%%%%%%%%%%%%%%
	%\begin{tocentry}
	
	%Some journals require a graphical entry for the Table of Contents.
	%This should be laid out ``print ready'' so that the sizing of the
	
	%Inside the \texttt{tocentry} environment, the font used is Helvetica
	%8\,pt, as required by \emph{Journal of the American Chemical
	
	%The surrounding frame is 9\,cm by 3.5\,cm, which is the maximum
	%permitted for  \emph{Journal of the American Chemical Society}
	%graphical table of content entries. The box will not resize if the
	%content is too big: instead it will overflow the edge of the box.
	
	%This box and the associated title will always be printed on a
	%separate page at the end of the document.
	
	%\end{tocentry}
	
	%%%%%%%%%%%%%%%%%%%%%%%%%%%%%%%%%%%%%%%%%%%%%%%%%%%%%%%%%%%%%%%%%%%%%
	
	%%%%%%%%%%%%%%%%%%%%%%%%%%%%%%%%%%%%%%%%%%%%%%%%%%%%%%%%%%%%%%%%%%%%%
	\begin{abstract}
		\textbf{Piezoelectric nanowires are highly attractive for the integration into flexible nanogenerators, capable of harvesting mechanical energy and to power small-scale wearable devices. The performance of such devices depends on both the piezoelectric and dielectric properties of the active region. In this work, we present a route to enhance the piezoelectric properties of group-III nitride nanowires by alloying them with transition metal nitrides. We first synthesize phase-pure wurtzite ternary \AlScN nanowires and integrate them into vertical nanogenerators. The resulting polymer-nanowire composite devices achieve effective piezoelectric charge coefficients of up to 8.5\,pC\,N$^{-1}$, largely exceeding those of AlN nanowires and thin films. Despite a reduced charge response compared to \AlScN thin films arising from the composite architecture, the concomitant decrease in dielectric permittivity compensates this effect, eventually yielding an enhanced voltage response. Effective medium modeling further indicates that additional performance improvements are anticipated from tailored device architectures. Our results establish \AlScN nanowires as a versatile platform for piezoelectric nanogenerators and provide general design guidelines for next-generation mechanical sensors and energy harvesting devices. 
		}

	\end{abstract}
	%%%%%%%%%%%%%%%%%%%%%%%%%%%%%%%%%%%%%%%%%%%%%%%%%%%%%%%%%%%%%%%%%%%%%
	%% Introduction
	%%%%%%%%%%%%%%%%%%%%%%%%%%%%%%%%%%%%%%%%%%%%%%%%%%%%%%%%%%%%%%%%%%%%%
	
	\section{Introduction}
	Alloying hexagonal wurtzite group-III nitrides with cubic rock-salt transition metal nitrides has led to new ternary compounds with exciting properties \cite{jena_jpn.j.appl.phys._2019}. The archetypal example is wurtzite \AlScN, which exhibits a piezoelectric charge coefficient ($\mathrm{d}_{33}$) up to five times higher than binary AlN for an optimum composition of $x\approx 0.35-0.40$ \cite{akiyama_adv.mater._2009,ambacher_j.appl.phys._2021,miyamoto_jpn.j.appl.phys._2025}. Moreover, ferroelectric switching of \AlScN thin films has been demonstrated in 2019 \cite{fichtner_j.appl.phys._2019}, thus marking the first observation of ferroelectricity in a group-III nitride semiconductor. These new functionalities in \AlScN emerge from lattice frustrations caused by the substitution of tetrahedrally coordinated Al atoms by Sc, which preferentially adopts an octahedral coordination \cite{urban_phys.rev.b_2021}. Although this mismatch renders wurtzite \AlScN metastable, it also opens new opportunities for device applications, including piezoelectric energy harvesting, which is currently dominated by lead-based compounds such as Pb(Zr,Ti)O$_{3}$ (PZT). 
	
	%%Fig.1
	%=======================================
	%AlScN NWs grown at different temperatures. Samples: (a) M2833, (b) M2839 (c) M2849, (d) from high to low T: M2833, M2835, M2839, M2848, M2849, M2853, M2871, (e) M2833+M2849. Results aquired by Adriano Notarangelo and Philipp John, with support from Hans Tornatzky in Raman measurements and discussions
	\begin{figure*}[b] 
		\centering
		\includegraphics[width=\textwidth]{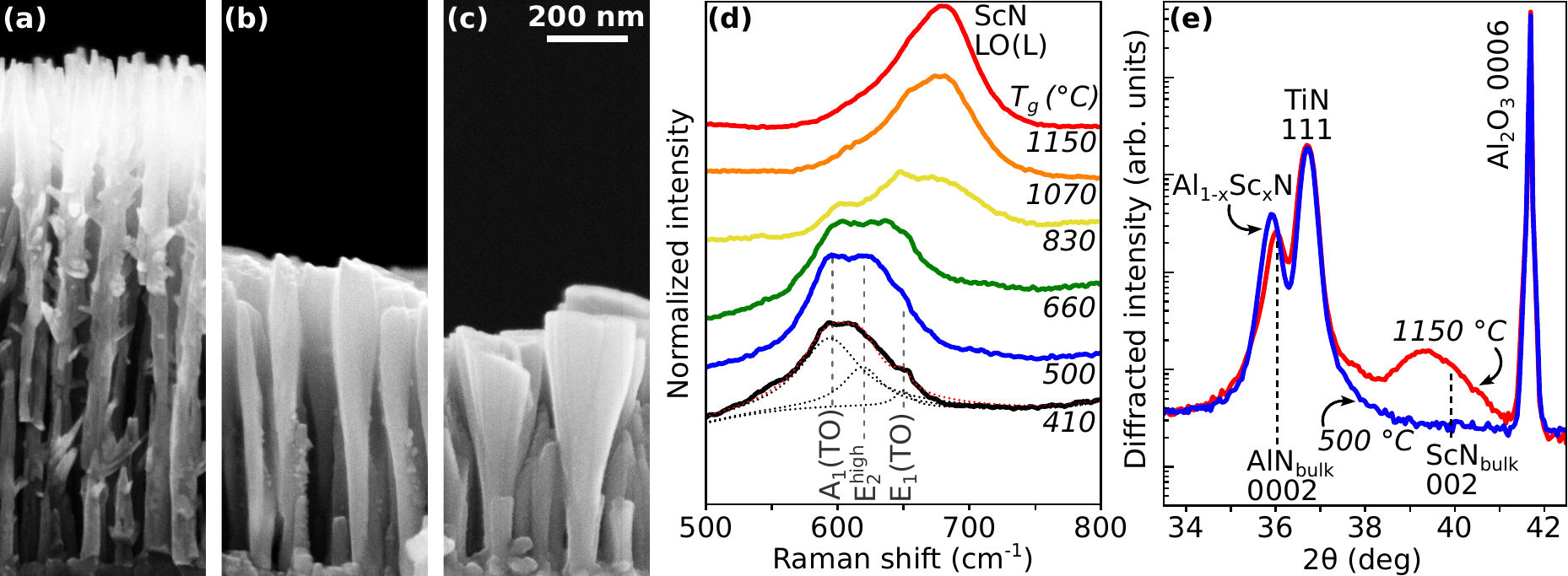}
		\caption{
			Cross-section secondary electron micrographs of Al$_{0.80}$Sc$_{0.20}$N segments grown on self-assembled AlN nanowire stems at (a) \SI{1150}{\degreeCelsius}, (b) \SI{830}{\degreeCelsius}, and (c) \SI{500}{\degreeCelsius}. (d) Room-temperature Raman spectra of several Al$_{0.80}$Sc$_{0.20}$N nanowire ensembles grown at different temperatures (T$\mathrm{_{g}}$), excited at \SI{473}{\nm}. Dashed lines indicate expected phonon positions of wurtzite Al$_{0.80}$Sc$_{0.20}$N \cite{solonenko_j.mater.sci._2020,solonenko_micromachines_2022}, dotted lines the corresponding Lorentzian fits. The peak at around 680\,cm$^{-1}$ is attributed to the LO(L) phonon of rock-salt ScN \cite{dinh_appl.phys.lett._2023,grumbel_phys.rev.mater._2024}. (e) Symmetric XRD 2$\text{$\theta$}$/$\mathrm{\omega}$ scans of the nanowire ensembles shown in (a) and (c). Dashed lines show the expected peak positions for bulk AlN(0002)\cite{nilsson_j.phys.appl.phys._2016} and ScN(002) \cite{niewa_chem.mater._2004}. %(bulk ScN 39.93$^\circ$, AlN  36.0350595$^\circ$)
		}
		\label{fig1}
	\end{figure*}
	
	%========================================================
	
	An attractive eco-friendly alternative to PZT-based piezoelectric devices are ZnO-based nanogenerators, first introduced in 2006 \cite{wang_science_2006}. A key advantage of these devices is their mechanical flexibility that is given by the use of piezoelectric nanowires instead of bulk ceramics. Furthermore, the nanowire geometry enables high-quality growth on dissimilar substrates by elastic strain relaxation \cite{zhang_cryst.growthdes._2011}, thereby offering improvements in both performance and device integration. However, ZnO nanowires suffer from screening effects due to unintentional \textit{n}-type doping, which limits their applicability for physiological sensing and energy harvesting at low frequencies. This doping typically arises from the inherent tendency of ZnO to incorporate native point defects (e.g., Zn interstitials and oxygen vacancies) or unintentional impurities, such as hydrogen \cite{ellmer_j.phys.appl.phys._2016}. One way to mitigate this issue is to replace ZnO by highly insulating AlN nanowires \cite{buatip_acsappl.nanomater._2024}. However, their maximum theoretical performance is inferior to that of ZnO, as reflected by the respective bulk piezoelectric coefficients ($\mathrm{d}_{33}^{\mathrm{AlN}}\approx5.4\,\mathrm{pC\,N^{-1}}$ and $\mathrm{d}_{33}^{\mathrm{ZnO}}\approx11.7\,\mathrm{pC\,N^{-1}}$) \cite{ambacher_j.phys.condens.matter_2002,lu_zincoxidebulkthinfilmsandnanostructures_2006}.

	Here, we aim to employ \AlScN nanowires in vertically integrated piezoelectric nanogenerators (VINGs) in order to overcome the piezoelectric performance limit of AlN. To our knowledge, there are three reports on the synthesis of such nanowires in the literature. \citet{bohnen_phys.statussolidia_2009} pioneered the growth of \AlScN nanowires by hydride vapor phase epitaxy in 2009, but achieved only a very low Sc incorporation of $x=0.05$. In contrast, \citet{zhang_crystengcomm_2024} sputtered \AlScN nanowires on TiN-buffered Si(111) substrates and achieved higher Sc contents ($0.08<x<0.17$), but with strong compositional gradients along the growth direction. Yet, their initial results point towards a strong enhancement of piezoelectric response upon Sc incorporation, as similarly reported for \AlScN thin films. Finally, \citet{wang_acsappl.nanomater._2024} grew $\approx$100\,nm-long \AlScN segments on self-assembled GaN nanowires by plasma-assisted molecular beam epitaxy (MBE) with a uniform Sc incorporation over a wide compositional range and achieved ferroelectric switching. The integration of \AlScN nanowires into functional nanogenerator devices remained yet undemonstrated.
	
	In this work, we grow $\approx$500\,nm long \AlScN segments on $\approx$200\,nm long AlN nanowire stems (referred to as \AlScN nanowires in the following). The AlN stems are self-assembled on 400\,nm-thick TiN/Al$_{2}$O$_{3}$(0001) templates as described in our previous works \cite{azadmand_phys.statussolidirrl-rapidres.lett._2020,auzelle_nanotechnology_2023,john_nanotechnology_2023}. We study growth over a wide temperature range, identify the optimum conditions to suppress phase separation, and achieve phase-pure wurtzite \AlScN nanowires over a large compositional range. Our self-assembly approach yields large-scale uniform nanowire ensembles and ensures simple device manufacturing by employing the metallic TiN layer as an integrated bottom electrode. This allows us to demonstrate, for the first time, large-area proof-of-principle nanogenerators based on \AlScN nanowires, for which we evaluate the  piezoelectric charge and voltage responses as well as the energy harvesting efficiency, and compare the results with bulk AlN and ZnO, and with commonly available \AlScN thin films.

	%% Results
	%%%%%%%%%%%%%%%%%%%%%%%%%%%%%%%%%%%%%%%%%%%%%%%%%%%%%%%%%%%%%%%%%%%%%  
	\section{Results and Discussion}
	\subsection{Epitaxy, morphology and phase of \AlScN nanowires}

	We first study the impact of growth temperature  during the plasma-assisted MBE of \AlScN nanowires on their morphology and crystalline phase. Figures\,\ref{fig1}a-c display representative cross-section secondary electron micrographs of  Al$_{0.80}$Sc$_{0.20}$N nanowires. Remarkably, decreasing the growth temperature from 1150 to 500\,$^\circ{}$C has drastic consequences for the ensemble morphologies. In particular, for \AlScN grown at the same temperature as the AlN stems (1150\,$^\circ{}$C, Figure\,\ref{fig1}a), long and thin nanowires form. Additionally, the high growth temperature induces extensive branching, with multiple branches nucleating and extending outward from the main body of each nanowire. In vapor-liquid-solid growth, this phenomenon usually occurs when catalyst droplets are intentionally deposited on the nanowire sidewalls, serving as new nucleation sites and giving rise to radial growth of branches \cite{cheng_nanotoday_2012}. For group-III nitride nanowires grown in absence of any catalyst, the self-induced branching observed here may result from phase separation of two structurally different materials, as similarly reported for In$_{2}$O$_{3}$/ZnO nanowires \cite{lao_nanolett._2002}. Notably, a decrease of growth temperature to 830\,$^\circ{}$C first reduces the formation of branches (Figure\,\ref{fig1}b), until it is completely suppressed at 500\,$^\circ{}$C (Figure\,\ref{fig1}c). In contrast, these ensembles are characterized by an inversely tapered nanowire morphology. This shape is commonly attributed to a limited diffusion length of group-III adatoms impinging on the nanowire sidewall facets, thereby favoring radial over axial growth and, eventually, reducing the overall average nanowire elongation rate \cite{daudin_nanotechnology_2021,john_nanotechnology_2023}.
	
	Next, we use Raman spectroscopy to determine the crystalline phase of our \AlScN nanowires. Figure\,\ref{fig1}d shows Raman spectra of ensembles grown in a wide temperature range, from 410 to 1150\,$^\circ{}$C. At low temperature the spectra are characterized by a broad band at around 610\,cm$^{-1}$, containing the contribution of several phonon modes. We attribute this series of peaks to phonon scattering in wurtzite \AlScN. In particular, the three most intense contributions at 594, 619 and 650\,cm$^{-1}$ in the ensemble grown at 500\,$^\circ{}$C fit well to the A$_{1}$(TO), E$_{2}^{\mathrm{high}}$, and E$_{1}$(TO) phonons reported for wurtzite Al$_{0.80}$Sc$_{0.20}$N, respectively \cite{solonenko_j.mater.sci._2020,solonenko_micromachines_2022}. Increasing the growth temperature further leads first to a decrease of scattering intensity of these phonons (660\,$^\circ{}$C), then to the appearance of a new band at around 680\,cm$^{-1}$ for growth temperatures $\geq$\,830\,$^\circ{}$C, eventually replacing the signal of wurtzite \AlScN. This peak coincides with the LO(L) phonon of rock-salt ScN  \cite{dinh_appl.phys.lett._2023,grumbel_phys.rev.mater._2024}, thus supporting the hypothesis of branching being a result of phase separation into wurtzite AlN and rock-salt ScN. The formation of rock-salt ScN at high substrate temperatures is attributed to the increased diffusivity of adatoms, allowing them to crystallize in their thermodynamically stable phases, namely wurtzite for AlN and rock-salt for ScN.

	We use x-ray diffraction (XRD) to investigate the phase separation in our samples in more detail. Figure\,\ref{fig1}e compares symmetric 2$\mathrm{\theta}$/$\mathrm{\omega}$ scans of the Al$_{0.80}$Sc$_{0.20}$N nanowire ensembles grown at 500 and 1150\,$^\circ{}$C. Besides the peaks at 36.77$^\circ{}$ and 41.69$^\circ{}$, corresponding to the TiN\,111 and Al$_{2}$O$_{3}$\,0006 reflections of the substrate, respectively, the diffractogram of the nanowires grown at 500\,$^\circ{}$C exhibits one single peak at around 35.92$^\circ$ that we attribute to diffraction from Al$_{0.80}$Sc$_{0.20}$N\,0002. The lower diffraction angle compared to bulk AlN\,0002 \cite{nilsson_j.phys.appl.phys._2016} evidences the incorporation of Sc that substitutes Al atoms in the wurtzite lattice. This substitution causes an increase in $c$-lattice constant as a function of Sc content for $0\leq x \leq 0.2$, as observed in previous theoretical and experimental reports on \AlScN thin films \cite{ambacher_j.appl.phys._2021, urban_phys.rev.b_2021}.

	The nanowire ensemble grown at 1150\,$^\circ{}$C, on the other hand, exhibits two peaks apart from those of the substrate. The peak at 36.02$^\circ$ coincides with the position of bulk AlN\,0002 and has a lower intensity than the \AlScN reflection of the ensemble grown at 500\,$^\circ{}$C. The broad peak at 39.48$^\circ$ is close to the position of bulk rock-salt ScN\,002,\cite{niewa_chem.mater._2004} clearly evidencing a separation of \AlScN into wurtzite AlN and nanocrystalline rock-salt ScN at high growth temperatures. We attribute the deviation of the ScN\,002 reflection towards lower angles compared to bulk ScN to the presence of strain at the AlN/ScN interface, resulting from the lattice mismatch between the two materials and/or a  mismatch in their thermal expansion coefficients. Note that the incorporation of Al into the rock-salt ScN phase would lead to peak shifts towards higher angles and can thus be excluded \cite{satoh_j.appl.phys._2022}. Importantly, an epitaxial out-of-plane orientation-relationship between the wurtzite and rock-salt phases is found (cf. Figure\,\ref{fig1}e), where AlN[0001]\,$\parallel$\,ScN[001]. 
	
	In order to investigate the in-plane orientation-relationship between the AlN and ScN in the phase-separated ensemble, we perform scanning transmission electron microscopy (STEM), as displayed in Figure\,S1 of the supporting information. Notably, Al- and Sc-rich domains are found to orient in vertical stripes along the nanowire axis, with the [10\={1}0] direction of the wurtzite AlN being parallel to the [110] direction of rock-salt ScN. This orientation-relationship of AlN[0001](10\={1}0)\,$\parallel$\,ScN[001](110) has been already observed between GaN and ScN during the growth of GaN/ScN core/shell nanowires \cite{john_nanolett._2024}, as well as during the thermal decomposition of \AlScN thin films \cite{hoglund_phys.rev.b_2010}. The branches (cf. Figure\,\ref{fig1}a), growing at an angle of (53 $\pm$ 5)$^\circ$ with respect to the nanowire axis, are likely a result of nucleation on the \{111\}-facets of rock-salt ScN, where an angle of $\approx$55$^\circ$ is expected \cite{john_nanolett._2024}.
	
	\subsection{Composition and microstructure of \AlScN nanowires}
	
	%%Fig.2
	%=======================================
	%AlScN NWs with different Sc content. Samples: EDX (from high to low Sc): M2887, M2883, M2877, M2874, M2847, M2900, M2901, measured by Aidan Campbell; RHEED M2874 NW stems (0% Sc), M2900(7% Sc), M2874(26% Sc), M2877 (32% Sc), M2883 (35% Sc), patterns taken by Adriano Notarangelo
	\begin{figure}[!ht]
		\includegraphics[width=\columnwidth]{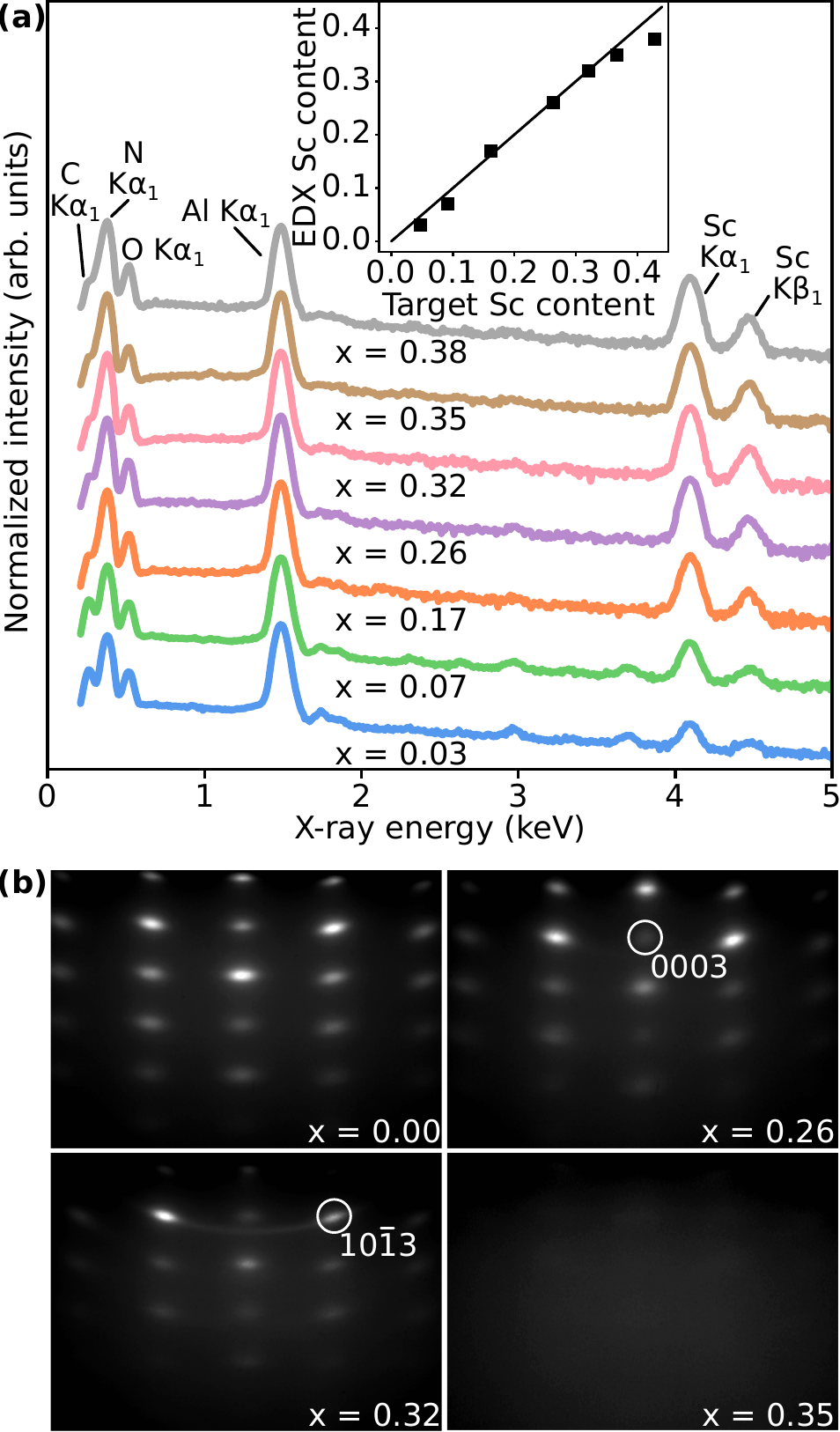}
		\caption{(a) Normalized SEM-EDX spectra of \AlScN nanowire ensembles with varying Sc content $x$, indicated below each spectrum. The inset shows the comparison with the target Sc content expected from the cell flux calibration. (b) RHEED patterns of selected ensembles (Sc content indicated in the bottom-right corner), recorded along the [11\={2}0] azimuth. The AlN RHEED pattern was acquired after the growth of the nanowire stems. The RHEED patterns of Al$_{0.74}$Sc$_{0.26}$N and Al$_{0.68}$Sc$_{0.32}$N were recorded at the end of growth, whereas and the pattern of Al$_{0.65}$Sc$_{0.35}$N after half of its growth duration. }
		\label{fig2}
	\end{figure}
	%=======================================================
	%Fig.3
	%=======================================
	%STEM investigation of M2874, aquired at CNRS-CRHEA by Ileana Florea and Philippe Vennegues, sample prep: Doreen Steffen, analysis: Philipp John
	\begin{figure*}[t] % Use [t] for top, [b] for bottom or [h] for here to control position
		\centering
		\begin{minipage}[c]{0.76\textwidth}
			\includegraphics[width=\textwidth]{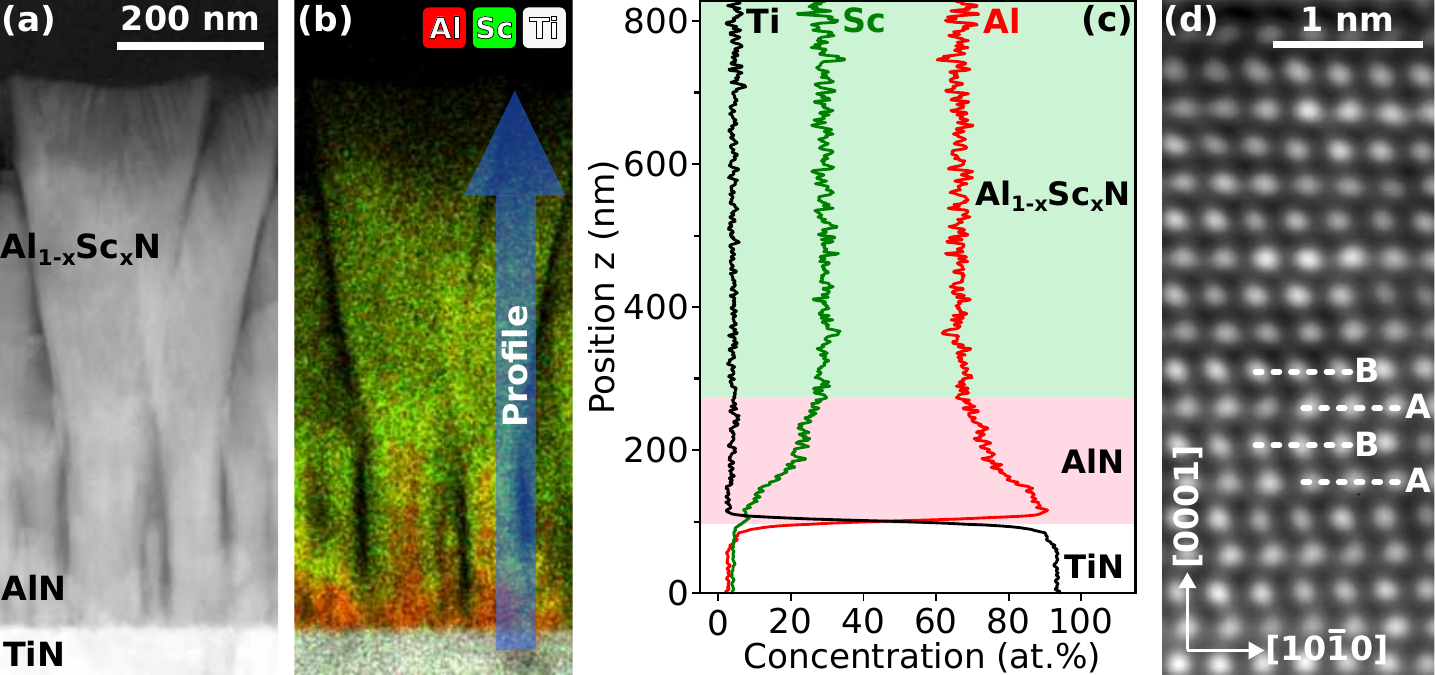}
		\end{minipage}%
		\hfill
		\begin{minipage}[c]{0.23\textwidth}
			\caption{
				Structural and chemical investigation of Al$_{0.74}$Sc$_{0.26}$N nanowires grown at \SI{410}{\degreeCelsius} by STEM. (a)~HAADF micrograph, (b) EDX map with (c) the corresponding EDX profile. (d)~Fourier filtered high-resolution micrograph, evidencing wurtzite phase.
			}
			\label{fig3}
		\end{minipage}
	\end{figure*}
	%========================================================
	
	After identifying the appropriate growth parameter window to stabilize the wurtzite phase in Al$_{0.80}$Sc$_{0.20}$N nanowires, we grow a sample series with varying Sc content $x$ on 2 inch wafers. To this end, we vary the Sc flux, keeping the Al- and active N-fluxes constant. The growth temperature is kept below 700\,$^\circ$C, and is further decreased with increasing Sc content in order to avoid any phase separation (see Table\,\ref{tab:growthconditions} in Section\,\ref{sec:experiments} for more details). Figure\,\ref{fig2}a shows normalized energy-dispersive x-ray spectroscopy (EDX) measurements of the different \AlScN nanowire ensembles recorded in a scanning electron microscope (SEM), which are used to determine their composition. The ratio of the Sc-related peak intensities (Sc\,K$\alpha$1 and Sc\,K$\beta$1) to the Al\,K$\alpha$1 peak progressively increases from the bottom to the top spectrum, evidencing an increased Sc incorporation. After quantification with the eZAF routine (cf. Section\,\ref{sec:experiments}), we obtain Sc contents between 0.03 and 0.38, covering essentially the entire accessible range for wurtzite \AlScN \cite{akiyama_adv.mater._2009,ambacher_j.appl.phys._2021,miyamoto_jpn.j.appl.phys._2025}. Moreover, the Sc content determined by EDX is in good agreement with the expected Sc content from the effusion cell flux calibration (cf. Table\,\ref{tab:growthconditions}), as shown in the inset of Figure\,\ref{fig2}a.
	
	To monitor the crystalline phase of the growing nanowires, we use reflection high-energy electron diffraction (RHEED). As displayed in Figure\,\ref{fig2}b, a transmission pattern is observed for the binary AlN nanowire stems, as expected for such group-III nitride nanostructures \cite{wolz_nanolett._2015,john_nanotechnology_2023}. The patterns of the \AlScN nanowires remain qualitatively unchanged until $x=0.26$, indicating that the nanowires crystallize in the wurtzite structure. However, a decrease in diffracted intensity is observed for the wurtzite 0003 double diffraction spot, which has been associated with increased mosaicity in \AlScN thin films. \cite{zhang_nanoscalehoriz._2023}. In the case of our nanowire ensembles, this could be explained by an increased tilt angle between individual nanowires.  Consistently, an increased intensity spread of the asymmetric Bragg reflections is found for the Al$_{0.68}$Sc$_{0.32}$N nanowire ensemble, being most prominent for the 10\={1}3 peak. In addition, faint circular features appear in the pattern of this ensemble, evidencing the formation of a polycrystalline phase. As the Sc content increases further ($x=0.35$), the RHEED pattern fades considerably, suggesting an amorphization that may occur as an intermediate phase towards the rock-salt phase of \AlScN, typically emerging at $x\geq 0.35$ \cite{satoh_j.appl.phys._2022}.
	Alternatively, the transition towards polycrystalline and amorphous phases for the nanowires with high Sc content may also be attributed to the well-known phenomenon of limited thickness epitaxy, occurring during the low-temperature MBE growth of various semiconductors \cite{eaglesham_j.appl.phys._1995}. This interpretation is consistent with the progressive decrease of crystalline quality observed \textit{in-situ} by RHEED during the nanowire growth of ensembles with $x> 0.30$ (data not shown), noting that our \AlScN segments are thicker than most other MBE-grown \AlScN layers, which typically exhibit thicknesses of around 25-200\,nm \cite{hardy_appl.phys.lett._2017,casamento_appl.phys.lett._2020,dinh_appl.phys.lett._2023a,nguyen_aplmater._2024}.
	
	In order to investigate the microstructure and chemistry of our wurtzite nanowires, we analyze the Al$_{0.74}$Sc$_{0.26}$N ensemble by STEM. The overview high-angle annular dark field (HAADF) micrograph in Figure\,\ref{fig3}a reveals a morphology characterized by inverse nanowire tapering, consistent with the other low-temperature ensemble shown in Figure\,\ref{fig1}c. EDX mapping reveals a homogeneous incorporation of Sc into the \AlScN segment grown on AlN nanowire stems (Figure\,\ref{fig3}b). The exctracted EDX profile (Figure\,\ref{fig3}c) confirms a composition of $x=0.28$, in good agreement with the value found by the measurements in Figure\,\ref{fig2}a. Moreover, strong compositional fluctuations are absent along the nanowire axis and perpendicular to it on the investigated length scale. Finally, high-resolution imaging within the \AlScN segment clearly evidences an \textit{AB} stacking sequence, characteristic for the desired wurtzite phase and in agreement with the RHEED analysis (Figure\,\ref{fig2}b). 
	
	%========================================================
	%%Fig.4
	%piezoelectric device fabrication and characterization done at CNRS-Neel by Rudeesun Songmuang, Natthawadi Buatip, Mostafa Saleh. Samples from low to high Sc-content: M2901, M2900, M2847, M2874, M2877, M2883. 
	
	\begin{figure*}[t] % Use [t] for top, [b] for bottom or [h] for here to control position
		\centering
		\begin{minipage}[c]{0.65\textwidth}
			\includegraphics[width=\textwidth]{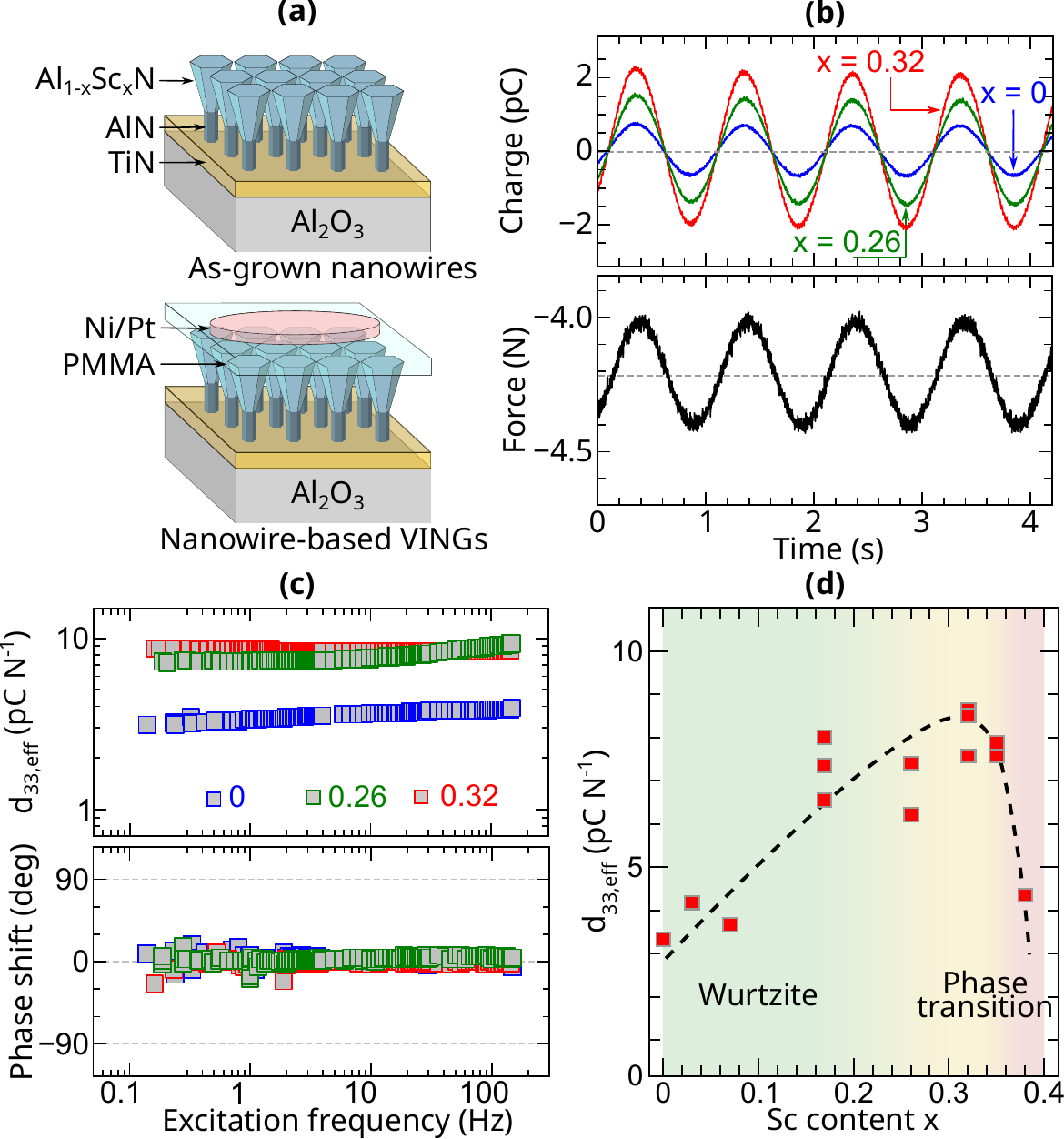}
		\end{minipage}%
		\hfill
		\begin{minipage}[c]{0.33\textwidth}
			\caption{
				(a) Schematic illustration of \AlScN-based VINGs (bottom), processed from the MBE-grown \AlScN/AlN nanowires (top), the TiN layer serving as bottom contact. (b) Charge response of \AlScN VINGs with varying Sc content $x$ to a sinusoidal force excitation at 1\,Hz.  (c) Extracted piezoelectric charge coefficient as a function of excitation frequency (upper panel), together with the phase shift between force and charge signals (lower panel). (d) Effective piezoelectric charge coefficient as a function of Sc content $x$ of the  \AlScN prototype VINGs. The dashed line serves as a guide to the eye.}
			\label{fig4}
		\end{minipage}
	\end{figure*}
	
	%============================================================================
	\subsection{\AlScN-based vertically integrated piezoelectric nanogenerators}
	
	To enable the use of our axial AlN/\AlScN nanowires grown on TiN/Al$_2$O$_3$ in functional piezoelectric devices, we process them into VINGs, as schematically illustrated in Figure\,\ref{fig4}a. First, a polymethyl methacrylate (PMMA) layer is spin-coated onto the nanowires to provide mechanical stability and to prevent short circuits during the subsequent electrode deposition. Owing to the high degree of coalescence, the PMMA forms a smooth, approximately 60\,nm-thick over-layer on top of the nanowires, with negligible penetration into the nanowire array. For the top contact we deposit Ni/Pt, while the metallic TiN layer serves as an integrated bottom contact. For each grown \AlScN ensemble, large-area devices with  a 0.8–1\,mm diameter top contact  are processed. The piezoelectric response of the VINGs is measured using a custom-built setup based on the Berlincourt method (i.e., in short-circuit conditions) \cite{npl2768}, while impedance measurements at the same position provide the dielectric loss in the frequency range of 20\,Hz to 1\,MHz \cite{buatip_acsappl.nanomater._2024}. Further details on device fabrication and measurement are provided in  Section\,\ref{sec:experiments}.

	Figure\,\ref{fig4}b displays the charge signal of selected \AlScN piezoelectric nanogenerators with Sc contents of $0$, $0.26$, and $0.32$ under a sinusoidal force excitation at 1\,Hz. A clear increase in amplitude is observed with increasing Sc content, reflecting enhanced piezoelectricity of the polymer-nanowire composite upon Sc incorporation.
	
	Next, we extract the effective piezoelectric charge coefficient d$_{33,\mathrm{eff}}$, corresponding to the ratio between the amplitude of the generated charge and that of the applied force. Figure\,\ref{fig4}c (top panel) shows d$_{33,\mathrm{eff}}$ as a function of excitation frequency in the range of 0.3--150\,Hz. The piezoelectric charge signal is almost constant in the entire frequency range, confirming the high mechanical and electrical stability of our nanogenerators. Furthermore, the phase shift between the force and charge signals (bottom panel) remains zero for all samples in the measured frequency range. Based on our previous results \cite{buatip_acsappl.nanomater._2024}, this indicates metal polarity of the \AlScN nanowires, consistent with the polarity of the AlN nanowire stems \cite{azadmand_phys.statussolidirrl-rapidres.lett._2020,jaloustre_acsappl.nanomater._2021,buatip_acsappl.nanomater._2024}.
	
	To evaluate the effect of Sc concentration on the performance of our \AlScN-based nanogenerators, Figure\,\ref{fig4}d compares d$_{33,\mathrm{eff}}$ as a function of Sc content of the processed devices, extracted at 1\,Hz. %Note that the \AlScN nanowire devices in this work exhibit a higher d$_{33,\mathrm{eff}}$ than the AlN nanowire composite devices reported in our previous study \cite{buatip_acsappl.nanomater._2024}, even at low Sc contents. This difference can be attributed to the device configuration, in particular the presence of a 2\,{\textmu}m-thick polydimethylsiloxane (PDMS) top layer in the earlier devices.
	A strong increase in d$_{33,\mathrm{eff}}$ by a factor of three is observed for the \AlScN VINGs compared to binary AlN nanowires, with a maximum of 8.5\,pC\,N$^{-1}$ at $x=0.32$. This value exceeds the piezoelectric charge coefficient of bulk AlN (5.4\,pC\,N$^{-1}$) \cite{ambacher_j.phys.condens.matter_2002} by nearly a factor of two and approaches that of bulk ZnO (11.7\,pC\,N$^{-1}$)\cite{lu_zincoxidebulkthinfilmsandnanostructures_2006}. The increase of d$_{33,\mathrm{eff}}$ as a function of Sc content has been attributed to a decrease in the $c$/$a$ lattice parameter ratio of the ternary alloy, as well as to lattice softening induced by the incorporation of Sc into the wurtzite lattice \cite{ambacher_j.appl.phys._2021,urban_phys.rev.b_2021}. For $x>0.35$, d$_{33,\mathrm{eff}}$ starts decreasing, most likely due to the amorphization observed in high--Sc content nanowire ensembles (cf. Figure \ref{fig2}b), which indicates the onset of the structural transition towards the non-piezoelectric rock-salt phase \cite{satoh_j.appl.phys._2022}.

	%========================================================
	%%Fig.5
	%effective medium modelling to  reproduce the trend of the AlScN NWs shown in Fig. 4. Determnation of dielectric constants done in CNRS-Neel by Rudeesun Songmuang, Natthawadi Buatip, Mostafa Saleh. Modelling established by Rudeesun Songmuang (CNRS-Néel) and Adriano Notarangelo (PDI)

	\begin{figure*}[t] % Use [t] for top, [b] for bottom or [h] for here to control position
		\centering
		\begin{minipage}[c]{0.65\textwidth}
			\includegraphics[width=\textwidth]{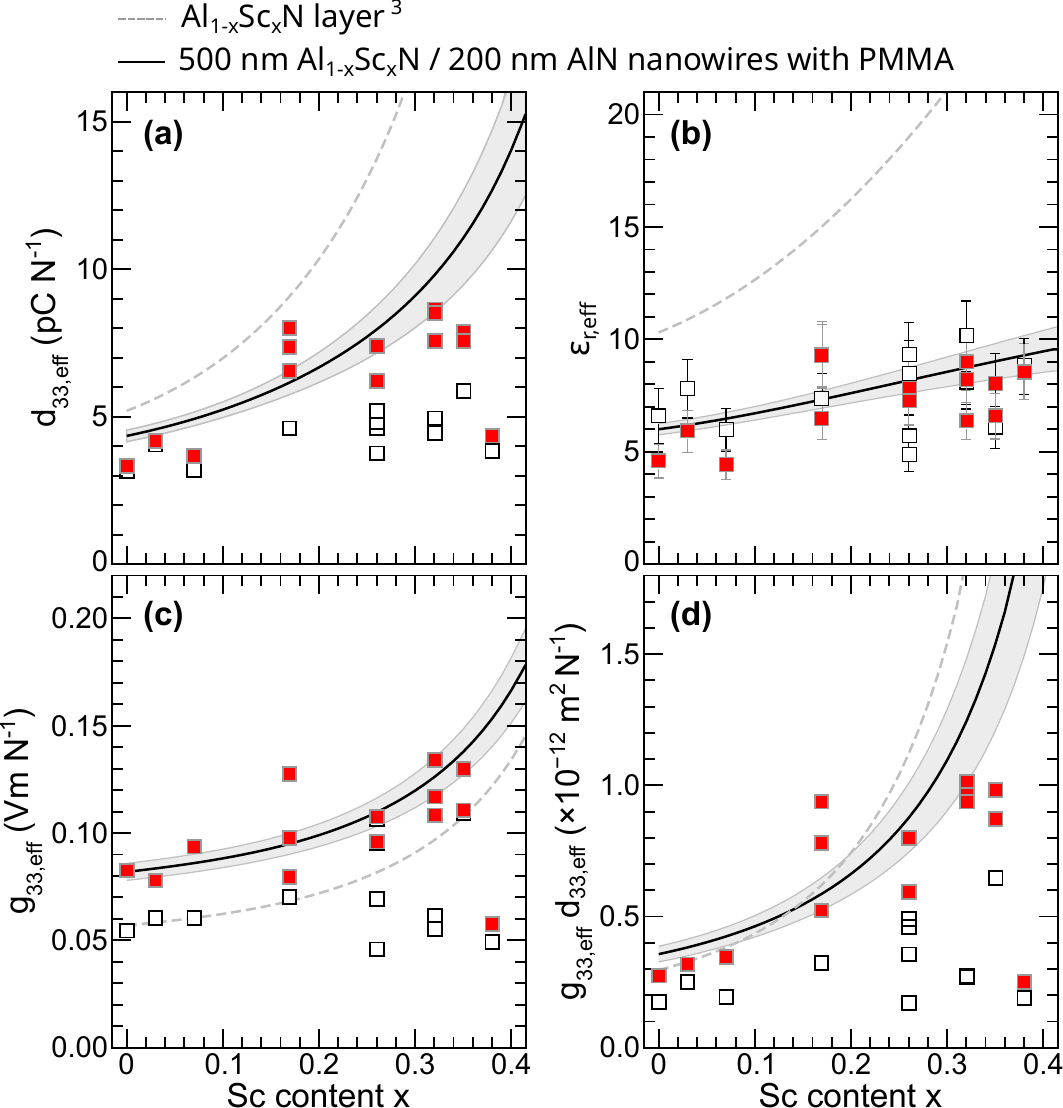}
		\end{minipage}%
		\hfill
		\begin{minipage}[c]{0.33\textwidth}
			\caption{
				Comparison of theoretical and experimental  effective piezoelectric and dielectric properties of the \AlScN VINGs. (a) Shows the piezoelectric charge coefficient, (b) the relative static permittivity, (c) the piezoelectric voltage coefficient, and (d) the energy harvesting figure of merit as a function of Sc content $x$. Devices exhibiting the strongest piezoelectric properties for a given composition are highlighted in red. The gray dashed line in each panel denotes the corresponding trend reported for \AlScN thin films \cite{ambacher_j.appl.phys._2021}. The black solid line indicates the calculation of each effective parameter as a function of Sc content, while the shaded region  represents the associated uncertainty.
			}
			\label{fig5}
		\end{minipage}
	\end{figure*}

	Although our \AlScN-nanowire based devices experience an increase of d$_{33,\mathrm{eff}}$ as a function of Sc content, the measured charge response remains lower than the values reported for \AlScN layers, which is expected due to the composite nature of our devices. 
	To interpret these results quantitatively, we apply a theoretical effective-medium approach and calculate the effective electromechanical properties of the nanowire-composite devices. Our model is based on analytical expressions derived by \citet{chan_ieeetrans.ultrason.ferroelectr.freq.control_1989} in the 1980s to describe the effective properties of micrometric 1-3 composites, i.e., consisting of one-dimensional piezoelectric rods embedded in a three-dimensional non-piezoelectric matrix. In this framework, the composite constituents are assumed to experience a uniform strain and electric field, corresponding to a parallel mechanical and electrical connectivity of the phases, analogous to a Voigt-type \cite{voigt_ann.phys._1889} averaging scheme. This approximation resembles the configuration of the \AlScN nanowire segments in our devices, which form a quasi 1–3 composite with air gaps between neighboring wires. We extend this approach by introducing an additional series connectivity to account for the vertical stacking of the \AlScN segments with the AlN stems and the PMMA top layer. In this case, the layers are mechanically and electrically connected in series, corresponding to a Reuss-type \cite{reuss_zamm-j.appl.math.mech.z.furangew.math.mech._1929} averaging scheme. The resulting model enables the calculation of the effective elastic, dielectric, and piezoelectric properties of the \AlScN VINGs from the intrinsic material parameters and volume fractions of the individual components in the nanowire–polymer composite structure. In particular, the modeling is carried out in three consecutive steps:
	
	(i) We first calculate  $d_{33,\mathrm{eff}}$, the effective elastic compliance $S_{33,\mathrm{eff}}$ and the effective dielectric permittivity $\varepsilon_{\mathrm{eff}}=\varepsilon_{0}\varepsilon_{r,\mathrm{eff}}$ of an \AlScN/AlN bilayer in series under \textit{iso-stress} conditions (i.e., identical internal pressure).  Here, $\varepsilon_{0}$ and $\varepsilon_{r,\mathrm{eff}}$ denote the vacuum permittivity and the effective relative permittivity, respectively. This calculation is done along the [0001] direction, using the Sc content dependent piezoelectric, elastic and dielectric properties of \AlScN and binary AlN published in the literature \cite{ambacher_j.appl.phys._2021}.
	(ii) Using the results from (i), we determine the effective properties of the active layer by forming a nanowire array in which the wires are arranged in parallel and separated by air gaps. This configuration corresponds to \textit{iso-strain} conditions (i.e., a common displacement) along the nanowire axis. 
	(iii) Finally, the response of the complete device is obtained by adding the non-piezoelectric PMMA layer onto the active layer from (ii) assuming \textit{iso-stress} conditions, from which $d_{33,\mathrm{eff}}$ and $\varepsilon_{\mathrm{eff}}$ of the full device are calculated.
	The exact formalism of the analytical model is provided in Section\,\ref{sec:modeling}.
	
	Figure\,\ref{fig5}a displays the results of our calculations as a function of Sc content, together with the experimentally obtained $d_{33,\mathrm{eff}}$ of \textit{all} processed VINGs for direct comparison. The white data points correspond to devices exhibiting a low response, which we attribute to challenges in prototype device processing and measurements, as well as to differences in nanowire length and composition between the center and edge of the 2 inch wafer.  Our calculations assume a heterostructure of 500\,nm-long \AlScN segments on top of 200\,nm-long AlN stems with a volume filling factor of 0.6, and a smooth, 60\,nm-thick PMMA encapsulation layer. The model (black solid line) shows good agreement with the trend observed in our best devices (red data points), which represent those primarily limited by the intrinsic properties of the \AlScN nanowires. Importantly, the lower $d_{33,\mathrm{eff}}$ compared to \AlScN layers (gray, dashed line) is solely explained by the series stacking of the \AlScN segments with the AlN stems and the non-piezoelectric PMMA top layer, i.e., to the device architecture. Note that the piezoelectric response of the \AlScN devices were calculated up to a Sc content of 0.5 assuming crystallization in the wurtzite phase. The experimental data points for $x>0.35$ are thus not well-reproduced by our analytical model.

	Although $d_{33,\mathrm{eff}}$ is essential to describe the piezoelectric properties of our nanowire-based devices, the key figure of merit for mechanical sensing applications is defined by the piezoelectric \textit{voltage} coefficient $g_{33,\mathrm{eff}} = d_{33,\mathrm{eff}} / \varepsilon_{\mathrm{eff}}$. To obtain $g_{33,\mathrm{eff}}$, we first experimentally extract $\varepsilon_{\mathrm{eff}}$ from impedance measurements performed at the same position used for the determination of $d_{33,\mathrm{eff}}$. In particular, $\varepsilon_{r,\mathrm{eff}}$ of each device is calculated from the measured capacitance and dielectric loss tangent. It is derived from the capacitance value measured at 200\,kHz, assuming a parallel-plate capacitor geometry and taking into account the device dimensions. At this frequency, the influence of leakage currents and low-frequency interfacial polarization is expected to be significantly reduced, as indicated by a low loss tangent (not shown). Consequently, the extracted $\varepsilon_{r,\mathrm{eff}}$ is close to the intrinsic dielectric response of \AlScN, which is predominantly governed by lattice contributions, while minimizing extrinsic effects. This approach enables a more reliable comparison of the influence of Sc incorporation on the dielectric properties. 
	
	Figure\,\ref{fig5}b shows the extracted $\varepsilon_{r,\mathrm{eff}}$ values as a function of Sc content. Due to the effective-medium nature of the nanowire-polymer composite,  $\varepsilon_{r,\mathrm{eff}}$ is strongly reduced compared to the dielectric permittivity of \AlScN layers (dashed, grey line). Our effective medium calculations (black, solid line) reproduce well this trend when considering the same device geometry used for the calculation of $d_{33,\mathrm{eff}}$ (cf. Figure\,\ref{fig5}a). 
	
	The determination of $\varepsilon_{r,\mathrm{eff}}$ enables the extraction of the corresponding $g_{33,\mathrm{eff}}$ values by combining the results shown in Figures\,\ref{fig5}a-b, which are presented in Figure\,\ref{fig5}c.  The experimental $g_{33,\mathrm{eff}}$ values of the best devices (red data points) clearly exceed those expected for \AlScN thin films (dashed, gray line), indicating that the reduction on $\varepsilon_{r,\mathrm{eff}}$ overcompensates the reduction of $d_{33,\mathrm{eff}}$. This is supported by our model predicting $\approx$1.5 times higher piezoelectric voltage coefficients of our \AlScN/AlN VINGs compared to \AlScN thin films. 
	
	Finally, we evaluate the figure of merit for energy harvesting of our devices, defined as $g_{33,\mathrm{eff}} d_{33,\mathrm{eff}}$, and compare it to the expected performance of \AlScN thin films. As shown in Figure\,\ref{fig5}d, a comparable overall efficiency is achieved experimentally and in the model, despite slightly lower output at high Sc contents.
	
	The effective-medium nature of our \AlScN nanowire-based VINGs, and the associated reduction in $\varepsilon_{r,\mathrm{eff}}$, has been shown to yield higher piezoelectric voltage coefficients and comparable  piezoelectric figures of merits to those of thin film counterparts. This is in agreement with the observations on the micrometric 1-3 piezoelectric composites studied by \citet{chan_ieeetrans.ultrason.ferroelectr.freq.control_1989}. These findings raise the question of whether the characteristics of VINGs can be further improved  by deliberately tailoring the device architecture or nanowire morphology. 
	From the perspective of device architecture, our model indicates that the  reduction of $d_{33,\mathrm{eff}}$ compared to \AlScN thin films is primarily attributed to the series coupling of the \AlScN segments with the less-piezoelectric AlN stems and the highly compliant, non-piezoelectric polymer. Replacing the AlN stems with conductive $n$-type GaN stems \cite{wang_acsappl.nanomater._2024}, for instance, is predicted to yield an additional increase of $\approx$10\,\% in $g_{33,\mathrm{eff}}$, and  $\approx$30\,\% in the product $g_{33,\mathrm{eff}} d_{33,\mathrm{eff}}$ in the Sc content regime around 0.3.
	Even more substential improvements can be achieved by engineering the nanowire morphology. In particular, reducing the nanowire fill factor leads to a linear decrase in $\varepsilon_{r,\mathrm{eff}}$, while leaving $d_{33,\mathrm{eff}}$ essentially unaffected \cite{chan_ieeetrans.ultrason.ferroelectr.freq.control_1989}. Since both $g_{33,\mathrm{eff}}$ and $g_{33,\mathrm{eff}} d_{33,\mathrm{eff}}$ scale inversely with $\varepsilon_{r,\mathrm{eff}}$, a further improvement by a factor of 3-4 is anticipated for a decrease of nanowire fill factor from 0.6 to 0.1, indicating a strong potential of nanowire-based devices for energy harvesting. Although such precise control is not achievable with our present self-assembly approach, it may be realized through large-area top-down fabrication techniques, as already demonstrated for GaN nanowires \cite{kang_nanotechnology_2024}.

	%In contrast, the filling factor of the nanowire array has negligible influence on $d_{33,\mathrm{eff}}$ (not shown), in agreement with previous findings \cite{chan_ieeetrans.ultrason.ferroelectr.freq.control_1989}. 
	%This observation is highly promising and suggests that further improvements in device architecture -- such as replacing AlN stems with conductive $n$-type GaN stems \cite{wang_acsappl.nanomater._2024} and depositing the contact or eliminating them altogether through large-area top-down processing \cite{kang_nanotechnology_2024} -- are expected to significantly enhance $d_{33,\mathrm{eff}}$, thus approaching the values reported for thin-film devices (dashed, grey line).
	%Furthermore, the use of top-down fabrication strategies yielding improved nanowire morphology would allow the encapsulation polymer to penetrate into the nanowire array, enabling -- beyond performance gains -- a straightforward exfoliation for use in flexible devices.
	
	%Again, even higher $g_{33,\mathrm{eff}}$ are anticipated for optimized device architectures combining an enhanced $d_{33,\mathrm{eff}}$ with similar dielectric properties.
	%%%%%%%%%%%%%%%%%%%%%%%%%%%%%%%%%%%%%%%%%%%%%%%%%%%%%%%%%%%%%%%%%%%%%
	%% conclusions
	%%%%%%%%%%%%%%%%%%%%%%%%%%%%%%%%%%%%%%%%%%%%%%%%%%%%%%%%%%%%%%%%%%%%%
	\section{Conclusion}
	In conclusion, we have investigated the growth of \AlScN nanowires by plasma-assisted MBE. Phase separation into wurtzite AlN and rock-salt ScN occurs at growth temperatures above 700\,$^\circ{}$C with an epitaxial orientation-relationship of AlN[0001](10\={1}0)\,$\parallel$\,ScN[001](110). Lower growth temperatures stabilize the desired wurtzite phase, with \AlScN nanowires being obtained in a wide compositional range ($0\leq x<0.35$). These nanowires are integrated into vertical piezoelectric nanogenerators, where the increased incorporation of Sc boosts the effective piezoelectric charge coefficients of our devices up to 8.5\,pC\,N$^{-1}$, clearly outperforming bulk AlN and approaching that of bulk ZnO. For Sc contents of $x>0.35$, a drop in $d_{33,\mathrm{eff}}$ occurs, explained by the amorphization of the material likely occurring before the phase transition towards the rock-salt \AlScN phase. Despite the lower charge signal of our devices compared to  \AlScN thin films, our nanogenerators exhibit an improved figure of merit for voltage sensing, and a comparable figure of merit for energy harvesting. This behavior is attributed to a lower effective dielectric permittivity in the nanowire-polymer composite. Effective medium modeling further reveals that additional performance gains can be achieved by optimizing the device architecture. In particular, a strongly piezoelectric active region with low dielectric permittivity, sandwiched directly between two electrodes, promises maximum response. Our model further demonstrates that the intrinsic advantages of nanowire architectures enable devices with performance levels beyond what is feasible with \AlScN thin films, thereby laying the groundwork for next-generation devices offering even higher efficiency and potentially flexiblity.

	%%%%%%%%%%%%%%%%%%%%%%%%%%%%%%%%%%%%%%%%%%%%%%%%%%%%%%%%%%%%%%%%%%%%%
	
	\section{Materials and Methods}
	
	\subsection{Experimental details}
	\label{sec:experiments}
	
	\AlScN is grown by plasma-assisted MBE on around \SI{200}{\nano\meter} long AlN nanowire stems, self-assembled on 400\,nm-thick metallic TiN films, the latter described in our previous works \cite{azadmand_phys.statussolidirrl-rapidres.lett._2020,john_nanotechnology_2023}. The AlN nanowire stems exhibit Al-polarity, as determined previously by  several methods \cite{azadmand_phys.statussolidirrl-rapidres.lett._2020,jaloustre_acsappl.nanomater._2021,buatip_acsappl.nanomater._2024}. To vary the composition of the growing \AlScN segments, we vary the Sc flux while keeping the Al and active N fluxes constant, as further detailed in Table\,\ref{tab:growthconditions}. The Al and Sc fluxes were calibrated by determining the growth rate of AlN and ScN layers in the Al- and Sc-limited regimes, respectively, while the active N flux was calibrated by determining the growth rate of GaN layers in the N-limited regime. All effusion cells are mounted with an angle of 38$^{\circ}$ with respect to the surface normal and growth is done under continuous sample rotation.  The growth temperature of the wurtzite \AlScN nanowire ensembles is kept below 700\,$^{\circ}$C and successively reduced as a function of Sc content in order to avoid phase separation and optimize for highest crystalline quality (cf. Table\,\ref{tab:growthconditions}). Substrate temperatures above 500\,$^{\circ}$C were measured via optical pyrometry at 920\,nm, calibrated according to a procedure we established previously \cite{azadmand_phys.statussolidirrl-rapidres.lett._2020,john_nanotechnology_2023}. For lower temperatures, a thermocouple positioned 10\,mm behind the heater filament is used. Low-temperature calibration is based on the melting point of In, occurring at 156.6\,$^{\circ}$C. Intermediate temperatures are estimated by a linear interpolation between pyrometric data and the In melting point.
	The crystallinity of the growing nanowires is monitored by RHEED, using an acceleration voltage of 20\,kV.
	
	\begin{table*}
		\caption{Growth conditions of \AlScN nanowires with different Sc content $x$.}
		\label{tab:growthparameters}
		%samples M2901, M2900, M2847, M2874, M2877, M2883, M2887.
		\begin{tabular}{c c c c c}
			\hline \hline
			Composition & Growth temperature  & N flux  & Al flux  & Sc flux \\
			&   ($^{\circ}$C) &   ($\times$10$^{15}$ at cm$^{-2}$s$^{-1}$) & ($\times$10$^{14}$ at cm$^{-2}$s$^{-1}$)  & ($\times$10$^{13}$ at cm$^{-2}$s$^{-1}$) \\
			
			\hline 
			Al$_{0.03}$Sc$_{0.97}$N & 660 & 1.5 & 2.2 & 1.1\\
			Al$_{0.07}$Sc$_{0.93}$N & 580 & 1.5 & 2.2 & 2.3\\
			Al$_{0.17}$Sc$_{0.83}$N & 500 & 1.5 & 2.2 & 4.3\\
			Al$_{0.26}$Sc$_{0.74}$N & 410 & 1.5 & 2.2 & 8.0\\
			Al$_{0.32}$Sc$_{0.68}$N & 410 & 1.5 & 2.2 & 10.5\\
			Al$_{0.35}$Sc$_{0.65}$N & 410 & 1.5 & 2.2 & 12.9\\
			Al$_{0.38}$Sc$_{0.62}$N & 240 & 1.5 & 2.2 & 16.7\\
			\hline \hline
		\end{tabular}
		\label{tab:growthconditions}
	\end{table*}
	
	SEM is used to investigate the nanowire morphology using a Hitachi S4800. The structural properties of the nanowires are characterized
	using a laboratory XRD system (Philips PANalytical X’Pert PRO MRD)
	equipped with a two-bounce hybrid monochromator Ge(220) for
	the Cu\,K$_{\alpha 1}$ source ($\lambda$\,=\,1.540598\,$\Angstrom$) and a \SI{1}{\milli\meter} receiving slit in front of the x-ray detector.
	The microstructure of selected samples was studied by STEM, using a Titan SPECTRA 200 (ThermoFisher), equipped with a cold field emission gun and a Cs aberration probe corrector. A dual energy-dispersive x-ray detector is used for EDX mapping. Cross-section specimens were prepared using standard mechanical grinding and dimpling methods, where final thinning was done using different acceleration voltages from 4 to 0.2\,kV with Ar ions, in a Gatan precision ion polishing system.

	The composition of all nanowire ensembles was determined using EDX spectroscopy (EDAX Octane Elect Super detector), exciting the sample with an electron beam of an energy of 8\,keV inside a dedicated SEM setup (Zeizz Ultra 55). To extract the atomic concentrations of Al and Sc, we used the well-separated Al\,K$_{\alpha 1}$ and Sc\,K$_{\alpha 1}$ peaks and analyze them with the ZAF algorithm of the EDAX software \cite{eggert_microsc.microanal._2021}, thus including corrections for the detector efficiency, atomic number, absorption and fluorescence of x-rays. To exclude any impact of the AlN stems, measurements were performed in the cross-section geometry, ensuring that the signal originated exclusively from the \AlScN segments.
	
	Raman spectroscopy was performed at room temperature, with a 473 nm laser line used for excitation. The spectra were acquired on the sample cross-section in a back scattering geometry, employing an Olympus microscope objective with a magnification of  $100\times$ and a numerical aperture of 0.9. A custom-modified HR Evolution setup (Horiba) was used, with a  800\,mm focal length monochromator equipped with a 1800\,grooves\,cm$^{-1}$ grating and a liquid nitrogen-cooled CCD detector. Spectral calibration was performed using the optical phonon of Si at 520.3\,cm$^{-1}$ as reference.

	To test the piezoelectric response of our \AlScN nanowires, we process VINGs. To this end, a 2\,\% PMMA solution is
	spin-coated at 6000\,rpm for 30\,s onto the \AlScN, forming a $\approx$60\,nm over-layer. After soft-baking at 180\,$^{\circ}$C for 5\,min, this encapsulation provided better mechanical stability and a more uniform surface for the subsequent contact fabrication. We use electron beam evaporation to deposit circular top contacts with a diameter of 0.8--1\,mm, consisting of 10\,nm Ni and 100\,nm Pt. Silver epoxy was then applied to the top contacts to improve the mechanical and electrical reliability. As bottom contact, we use the 400\,nm-thick sputtered TiN template. The direct piezoelectric response of the nanogenerators is measured according to the Berlincourt principle \cite{npl2768} and applying a dynamic (0.3 to 150\,Hz) sinusoidal force of 0.2\,N, with a static preload of 3--4\,N to ensure device stability. After measuring the piezoelectric signal, an impedance measurement was performed at the same position under the same static preload using an LCR meter (Keysight 4980AL) with an excitation voltage of 200\,mV and frequency ranging from 20\,Hz to 1\,MHz. The capacitance and dielectric tangent loss values of the devices under test were measured.
	
	Large language models have been used to improve the written language, clarity and flow of the manuscript. 
	
	\subsection{Effective medium modeling}
	\label{sec:modeling}

	%========================================================
	
	%Averaging models schematics
	\begin{figure}
		\includegraphics[width=\columnwidth]{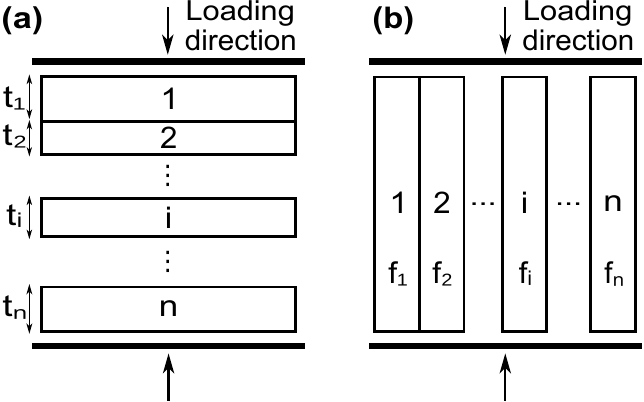}
		\caption{Schematic representation of a sequence of layers connected (a) in series  and (b) in parallel  with respect to the applied stress.}
		\label{fig6}
	\end{figure}
	%============================================================================
	
	To evaluate the effective dielectric, elastic and piezoelectric properties of our \AlScN/AlN nanowire-polymer composites subjected to mechanical loading along the nanowire axis, we employed an effective medium approach. 
	To derive expressions describing the effective properties, we begin with the fundamental constitutive equations of piezoelectricity for both the direct and inverse effects and apply the appropriate boundary conditions. In the experimental configuration of the Berlincourt method, the measurement is performed under short-circuit conditions, i.e., the external voltage across the sample is zero. Consequently, the total voltage across the composite vanishes, whereas the local electric field within each constituent may in general be non-zero. The remaining boundary conditions are determined by the specific geometrical arrangement of the composite phases, as detailed below.
	
	\paragraph{Series electrical connection}
	In a series of stacked layers (see Figure\,\ref{fig6}a), an iso-stress condition throughout the layers can be assumed, corresponding to a Reuss  averaging scheme \cite{reuss_zamm-j.appl.math.mech.z.furangew.math.mech._1929}. Furthermore, for such a layer stack sandwiched between two electrodes, the electric displacement is continuous across interfaces, and the voltage divides according to the permittivity of each layer. Under these boundary conditions, the effective piezoelectric charge coefficient of the stack is given by a thickness- and permittivity-weighted average of the individual layers 
	
	\begin{equation}
		d_{33,\mathrm{eff}}^{\mathrm{series}}=\frac{\sum_{i} t_{i}\frac{d_{33,i}}{\varepsilon_{i}}}{\sum_{i} \frac{t_{i}}{\varepsilon_{i}}},
	\end{equation}
	
	\noindent where $t_{i}$, $\varepsilon_{i}$, and $d_{33,i}$ are the thickness, absolute permittivity ($\varepsilon_i=\varepsilon_{r,i}\varepsilon_0$) and piezoelectric charge coefficient of the i$^{\mathrm{th}}$ layer, respectively. Under the same electrical and mechanical boundary conditions, the effective elastic compliance of the stack is calculated by
	
	\begin{equation}
		S_{33,\mathrm{eff}}^{\mathrm{series}}=\frac{\sum_{i} t_{i}\left(S_{i}-\frac{d_{33,i}^{2}}{\varepsilon_{i}}\right)}{t_{\mathrm{tot}}}+\frac{\left(\sum_{i} \frac{t_{i}d_{33,i}}{\varepsilon_{i}}\right)^{2}}{t_{\mathrm{tot}}\sum_{i} \frac{t_{i}}{\varepsilon_{i}}},
	\end{equation}
	
	\noindent where $S_{i}$ denotes the elastic compliance of the  i$^{\mathrm{th}}$ layer and $t_{\mathrm{tot}}$ the total thickness of the layer stack. Finally, the effective permittivity of such a series layer stack is given by
	
	\begin{equation}
		\varepsilon_{33,\mathrm{eff}}^{\mathrm{series}}=\left(\sum_{i} \frac{ \frac{t_{i}}{t_{\mathrm{tot}}}}{\varepsilon_{i}}\right)^{-1},
	\end{equation}

	\paragraph{Parallel electrical connection}
	For phases connected in parallel (see Figure\,\ref{fig6}b), an iso-strain condition can be assumed, corresponding to a Voigt  averaging scheme \cite{voigt_ann.phys._1889}. In this configuration, the electric field remains uniform across all constituents. The effective piezoelectric charge coefficient of such a composite is therefore given by
	
	\begin{equation}
		d_{33,\mathrm{eff}}^{\mathrm{par}}=\frac{\sum_{i} f_{i}\frac{d_{33,i}}{S_{i}}}{\sum_{i} \frac{f_{i}}{S_{i}}},
	\end{equation}
	
	\noindent where $f_{i}$ denotes the volume fraction  of each constituent. Here, the weighting by $1/S_{i}$ reflects the iso-strain boundary condition, under which all phases deform equally along the polarization direction, while the internal stress redistributes according to their stiffness.  The effective elastic compliance in this parallel  geometry is given by
	
	\begin{equation}
		S_{33,\mathrm{eff}}^{\mathrm{par}}=\frac{1}{\sum_{i} \frac{f_{i}}{S_{i}}},
	\end{equation}
	
	\noindent while the effective  permittivity  is calculated by
	
	\begin{equation}
		\varepsilon_{33,\mathrm{eff}}^{\mathrm{par}}=\sum_{i} f_{i}\varepsilon_{i}.
	\end{equation}

	%%%%%%%%%%%%%%%%%%%%%%%%%%%%%%%%%%%%%%%%%%%%%%%%%%%%%%%%%%%%%%%%%%%%%	

	%\begin{suppinfo}
	%		
	%		% A listing of the contents of each file supplied as Supporting Information
	%		% should be included. For instructions on what should be included in the
	%		% Supporting Information as well as how to prepare this material for
	%		% publications, refer to the journal's Instructions for Authors.
	%		% 
	%		The following files are available free of charge.
	%		\begin{itemize}
		%			\item Filename: brief description
		%		\end{itemize}
	%		% 
	%	\end{suppinfo}

\begin{acknowledgement}
	The authors thank their PDI colleagues D. Steffen for TEM sample preparation, N. Volkmer for assistance with SEM measurements, and C. Stemmler for dedicated MBE maintenance. They further acknowledge D. Mornex from Néel's SERAS platform for technical assistance in the development of the piezoelectric measurement setup. We are further grateful to M. Yuan for critically reading the manuscript. This work was funded by Deutsche Forschungsgemeinschaft (490935200) and Agence Nationale de la Recherche (ANR-21-CE09-0044) through the project Nanoflex, and supported by COST Action OPERA from the European Network for Innovative and Advanced Epitaxy, CA20116 (www.cost.eu).
\end{acknowledgement}



\bibliography{AlScN}

\end{document}